\documentclass{nature}
\usepackage{graphicx}

\usepackage{color}
\usepackage{amsmath}
\usepackage{amssymb}
\usepackage{graphics}
\usepackage{epsfig}
\usepackage{upgreek}
\usepackage{lineno}
\usepackage{dcolumn}
\usepackage{bm}
\usepackage[utf8]{inputenc}
\usepackage[T1]{fontenc}
\usepackage{booktabs, array, mathptmx, float, tabularx, booktabs, lipsum, amsmath,multirow}
\usepackage{siunitx, xcolor}
\usepackage[version=4]{mhchem}
\usepackage{color}
\usepackage{subfigure}
\usepackage[colorlinks,linkcolor=blue,anchorcolor=blue,citecolor=blue]{hyperref}

\title{Directional intense terahertz radiation driven by abruptly autofocusing lasers in air}
\author{Xiao-Ran Zheng,$^{1,\dag}$ Nan Li,$^{2,3,\dag}$ Wei-Min Wang,$^{2,3,4,*}$ Rui Zhang,$^{4}$ Cun-Lin Zhang,$^{1}$ Liang-Liang Zhang$^{1,*}$}

\begin{document}

\maketitle

\begin{affiliations}
\item Key Laboratory of Terahertz Optoelectronics (Ministry of Education), Department of Physics, Capital Normal University, Beijing 100048, China
\item Department of Physics and Beijing Key Laboratory of Opto-electronic Functional Materials and Micro-nano Devices, Renmin University of China, Beijing 100872, China
\item Key Laboratory of Quantum State Construction and Manipulation (Ministry of Education), Renmin University of China, Beijing 100872, China
\item IFSA Collaborative Innovation Center, Shanghai Jiao Tong University, Shanghai 200240, China
\item Shenzhen Institute of Advanced Technology, Chinese Academy of Sciences, Shenzhen 518055, China

\end{affiliations}

\leftline{$^\dag$These authors contributed equally to this work.}
\leftline{$^*$ Corresponding authors}
\leftline{W.-M.W. (email: weiminwang1@ruc.edu.cn), L.-L.Z. (email: liangliang\_zhang@cnu.edu.cn)}\vspace*{1cm}

\begin{abstract}
Two-color laser induced plasma filamentation in air could serve as tabletop sources of broadband terahertz (THz) pulses. Ubiquitous air in the earth facilitates its widespread utilities, particularly, in wireless communication and remote sensing, exploiting the unique advantage of the air-based scheme that the THz sources can be delivered over standoff distances via the pump laser propagation in air. However, the THz emission pattern inevitably has a conical angular profile with a dip in the propagation axis, therefore, THz energy concentration, propagation directionality, and accuracy of signal demodulation are significantly impaired, greatly limiting its direct applications. Here, we successfully eliminate the unfavorable conical profile by experiments and meanwhile enhance THz directionality and intensity by 17 folds by use of abruptly-autofocusing laser beams. Our theory and simulations show that these observations are attributed to efficient suppression of the dephasing effect appearing in previous investigations with ordinary Gaussian laser beams. This scheme is easily accessible since the abruptly-autofocusing beam can be achieved by imposing a spatial light modulator on the input Gaussian beam. This study solves a long-standing problem in the two-color laser scheme and clarifies the underlying physics of the conical angular profile formation.
\end{abstract}

Achieving intense tabletop terahertz (THz) pulse sources with high energies, broad bandwidths and favorable spatial profiles is crucial in THz science and technology, which paves the way toward the THz applications in nonlinear optics, biomedical spectroscopy, security inspection, communication, etc \cite{ref1,ref2,ref3,ref4}. Currently, optical rectification in electro-optic crystal \cite{ref5,ref6} and two-color laser induced filamentation in air \cite{ref7,ref8} are two major methods for intense THz pulse generation. By optical rectification, high-energy THz pulses with Gaussian spatial profiles can be obtained. However, the optical damage threshold of electro-optic crystal limits further increase of the THz energy and the THz spectrum bandwidth is usually lower than 5 THz \cite{ref9}. From two-color laser induced plasma filamentation in air, the generated THz pulses have ultra-broad bandwidths exceeding 30 THz and field strengths reaching 100 MV/cm \cite{ref10}, in principle which can be further increased since air plasma has nearly no optical damage threshold. In addition to these benefits in the air-based scheme, ubiquitous air in the earth facilitates its widespread use for various application scenarios. Particularly, this air-based scheme provides an efficient approach to deliver the THz pulse sources over standoff distances via the pump laser propagation in air, which has unique advantages in THz communication and remote sensing \cite{ref11} over other optical techniques.

Nonetheless, the generated THz pulse usually has a conical angular profile with a dip in the propagation axis \cite{ref12,ref13,ref14}, especially at high frequencies \cite{ref15,ref16}. Most previous studies focused on enhancing the THz generation efficiency, such as changing gas media \cite{ref17,ref18}, lengthening laser wavelength \cite{ref19}, and adopting multi-color laser excitation \cite{ref20}, but the conical angular profile with a dip of the generated THz pluses inevitably appears. Such an angular profile with peak intensity deviating from the center leads to poor performance in propagation, collection, and beam profile maintenance compared to other THz sources such as electro-optic crystal and photoconductive antenna with Gaussian profiles. For instance, the conical angular profile significantly impairs energy concentration, propagation directionality, and accuracy of signal demodulation demanded in THz communication \cite{ref4}, remote sensing \cite{ref11}, and large-area imaging \cite{ref21}. Thus, it is highly expected to overcome the THz conical angular profile and achieve a Gaussian profile.

This conical angular profile is considered to be caused by the relative phase shift between the two-color laser beams as they propagate in the air plasma filament \cite{ref22,ref23,ref24,ref25}. The relative phase determines the strength and polarity of the net current, therefore, affects the intensity and angular distribution of the emitted THz pulse. The phase shift (or dephasing effect) can reduce the THz intensity in the propagation axis and form a dip in the THz angular distribution, particularly with the dephasing length smaller than the filament length \cite{ref22,ref23,ref24,ref25}. In this work, our experiments and far-field simulations show that the dip cannot be eliminated even when the filament length is only a few mm, shorter than the usually-adopted dephasing length $\sim$20 mm. We find that this dephasing length is significantly overestimated because the Gouy phase of Gaussian beams is ignored. With the Gouy phase, the dephasing length drastically reduces to $\sim$0.6 mm and hence the dip should still exist, which can well explain our observations as well as the previous experimental results \cite{ref13,ref16,ref22,ref23,ref24}.

Here, we propose to use two-color abruptly-autofocusing (AAF) beams to eliminate the unfavorable THz conical profile with a dip. We convert ordinary Gaussian beams to AAF ones by a spatial light modulator (SLM) with a transverse phase distribution. This phase acts as a “multi-focal-length lens” and causes the dephasing length to increase in the two-color laser focusing range. Our calculation shows that the dephasing length increases by 5-fold up to $\sim$3 mm, the dephasing effect is overcome, and then the conical profile can be eliminated as well as the THz emission directionality can be improved. These agree with our experimental observations. Due to the suppressed dephasing effect, more THz source elements emitted from the filament have the same polarity, and coherent superposition of these source elements can considerably enhance the overall THz intensity. Therefore, this scheme provides a simple solution to the long-standing problem of the unfavorable THz conical angular profile and further strengthens the THz emission. Note that THz energy, spectrum, and polarization regulated with AAF beams were reported \cite{ref26,ref27,ref28}, but influences on THz angular profile have not been studied yet.

\begin{figure}[!htbp]
	\centering{\epsfig{figure=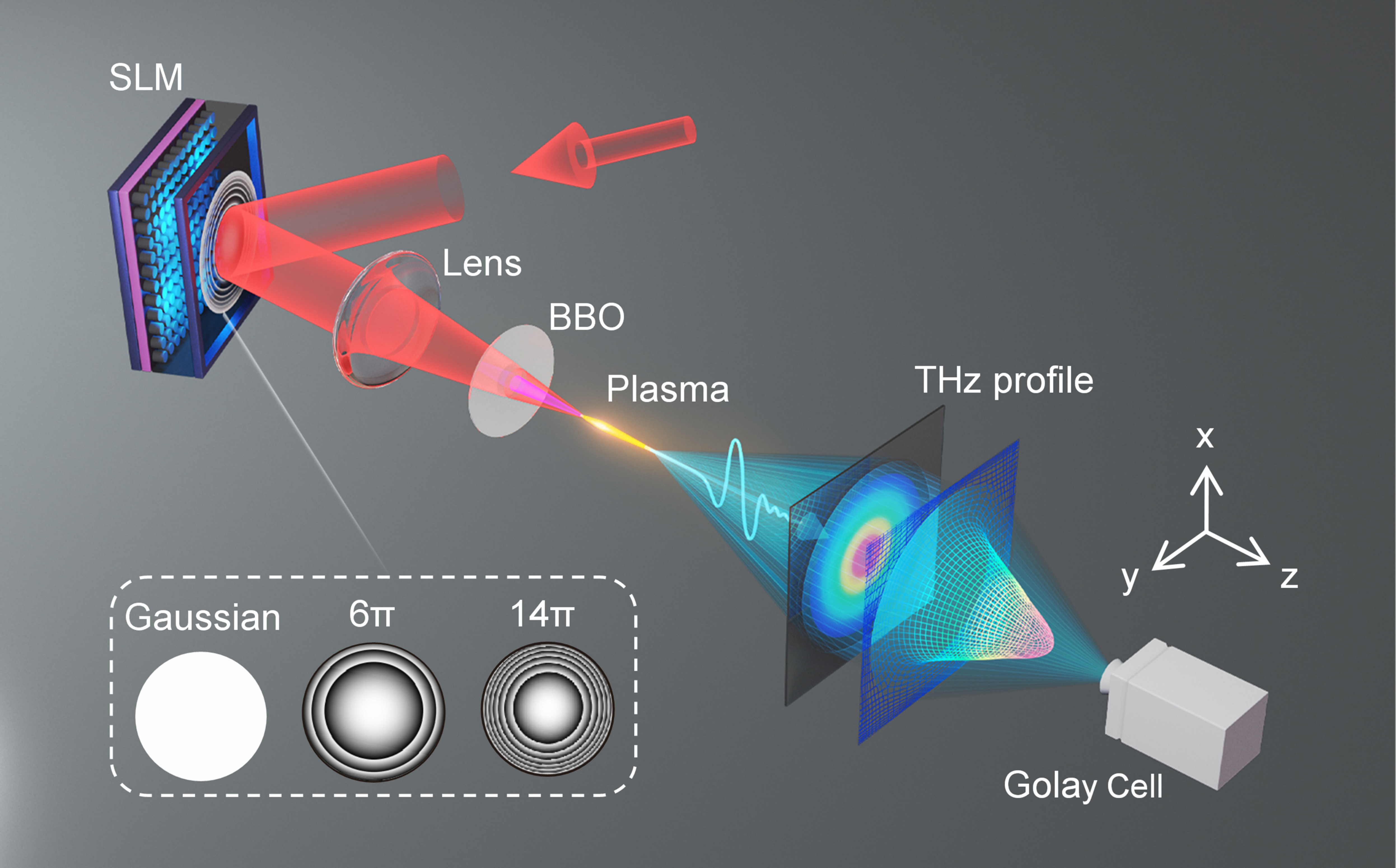, width=1.0\columnwidth}}
	\caption{\textbf{$|$ Schematic diagram of the experimental setup.} SLM: spatial light modulator, BBO: $\beta$-barium borate crystal. The inset illustrates the phase mask distributions for a Gaussian beam and AAF beams with the modulation depths of $6\pi$ and $14\pi$ as examples.\label{Fig1}}
\end{figure}

\textbf{Directional THz emission without a dip observed in experiments.}
Schematic of the experimental setup is shown in Fig. \ref{Fig1}. Gaussian laser beams with adjustable wavelength from 1200 nm to 1600 nm (1 KHz repetition rate) are delivered by an optical parametric amplifier (OPA, Spectra Physics). To achieve the best modulation efficiency, the laser beam is expanded to fully cover the liquid crystal chip of the spatial light modulator (SLM, Hamamatsu LCOS-X15213-08). The phase mask exerted on SLM is loaded with an additional transverse spatial phase  $\psi_{SLM}\left(R\right)=-CR^{3}/w_{SLM}^{3}$, where $C$ is the modulation depth of additional transverse spatial phase, $w_{SLM}$ is the radius of optical laser beam irradiated on the SLM, and $R$ is the transverse spatial position. With a plain phase mask, i.e., $C=0$, the output beam is still a Gaussian one, as illustrated in Fig. \ref{Fig1}. When $C$ increases to $2\pi\text{-}16\pi$ in our experiments, the output beam turns to an AAF one. Then, an air plasma filament is formed by focusing the output laser beam together with its second harmonic generated through a 100-$\mu$m-thick type-I $\beta$-barium borate (BBO) crystal by means of a plano-convex lens with 100 mm focal length. Fluorescence image of the two-color plasma filament is recorded using a charge coupled device (CCD) camera. The initial relative phase and intensity proportion of the two-color laser beams are optimized to most efficiently generate THz emission by adjusting the position and angle of the BBO crystal. Forward THz emission is collected after eliminating the residual laser beams by a silicon wafer and low-pass filters (QMC Instruments Ltd). To characterize the THz emission spatial distribution, two-dimensional raster scanning is performed using a Golay cell (Microtech SN: 220712-D) on the cross-sectional plane 6cm away from the plasma filament end in the forward direction. The THz time-domain waveform is detected by electro-optical sampling through a 1-mm-thick ZnTe crystal, as shown in Fig. S1 in Supplementary Materials (SM).

\begin{figure}[!htbp]
	\centering{\epsfig{figure=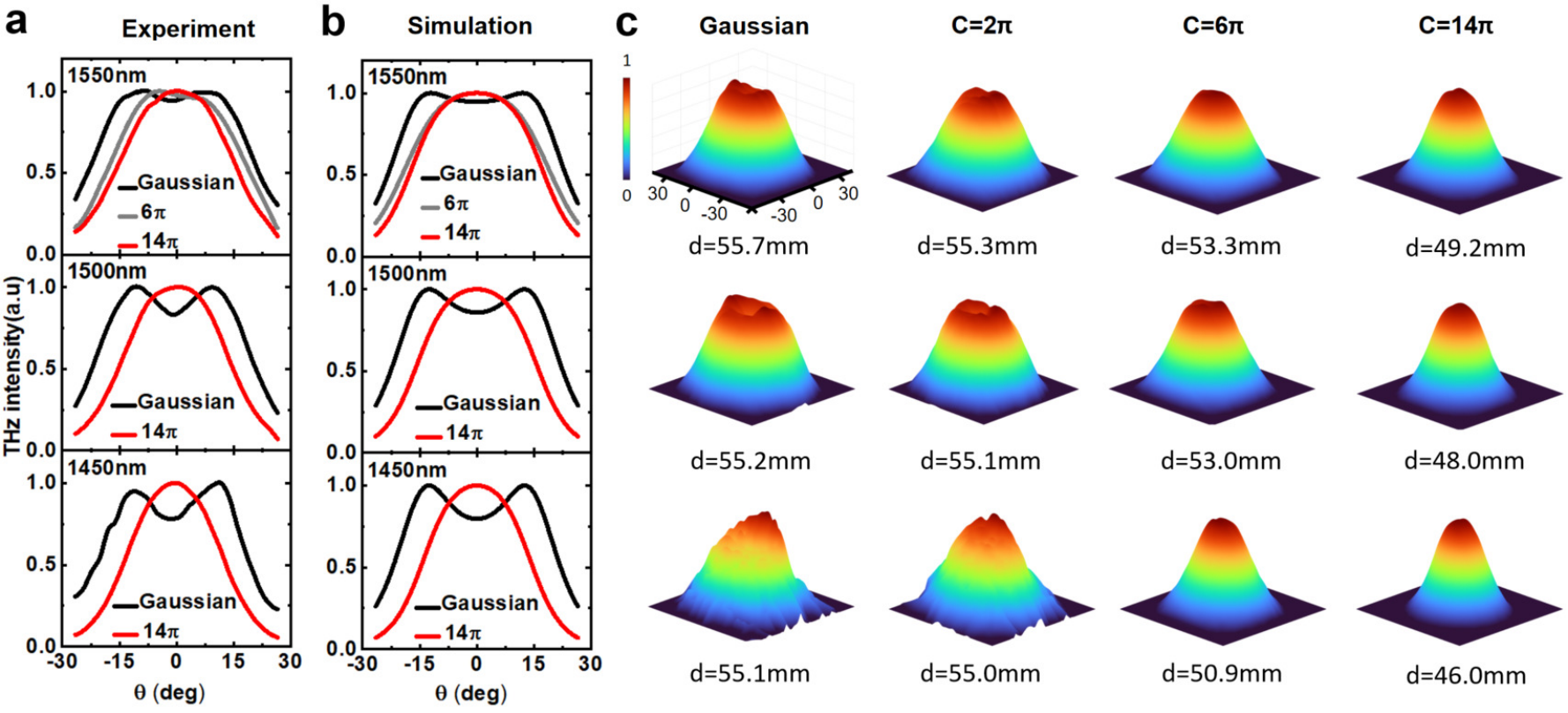, width=1.0\columnwidth}}
	\caption{\textbf{$|$ Angular distributions of THz pulses.} \textbf{a}, Experimentally measured and \textbf{b}, theoretically simulated angular distributions of THz pulses generated by the Gaussian beams (black line) and AAF beams with $C=6\pi$ (gray line) and $14\pi$ (red line), respectively, where different laser wavelengths are taken in each column and far-field simulations [25] are performed. THz intensity is normalized by its maximum in each curve. \textbf{c}, Cross-sectional normalized intensity profiles of the THz pulses generated by the Gaussian beams and AAF beams with $C=2\pi$, $6\pi$, and $14\pi$, respectively, where different laser wavelengths are taken in each row and the corresponding THz diameter $d$  is shown on the bottom of each plot.   \label{Fig2}}
\end{figure}

THz angular distributions measured in our experiments are shown in Fig. \ref{Fig2}a. A dip in the THz propagation axis is observed when the Gaussian laser beams (black line) are adopted, as reported in most existing experimental investigations \cite{ref13,ref22,ref23}. The dip becomes deeper as the fundamental-laser wavelength decreases from 1550 nm to1450 nm, where the second-harmonic laser wavelength correspondingly decreases. By contrast, this dip completely disappears when the AAF beam with the modulation depth $C=14\pi$ is used (red line) for each wavelength. As $C$ increases from $2\pi$ to $14\pi$, the dip fades, accompanying with less divergence angel or better emission directionality [see Fig. \ref{Fig2}c].

\begin{figure}[!htbp]
	\centering{\epsfig{figure=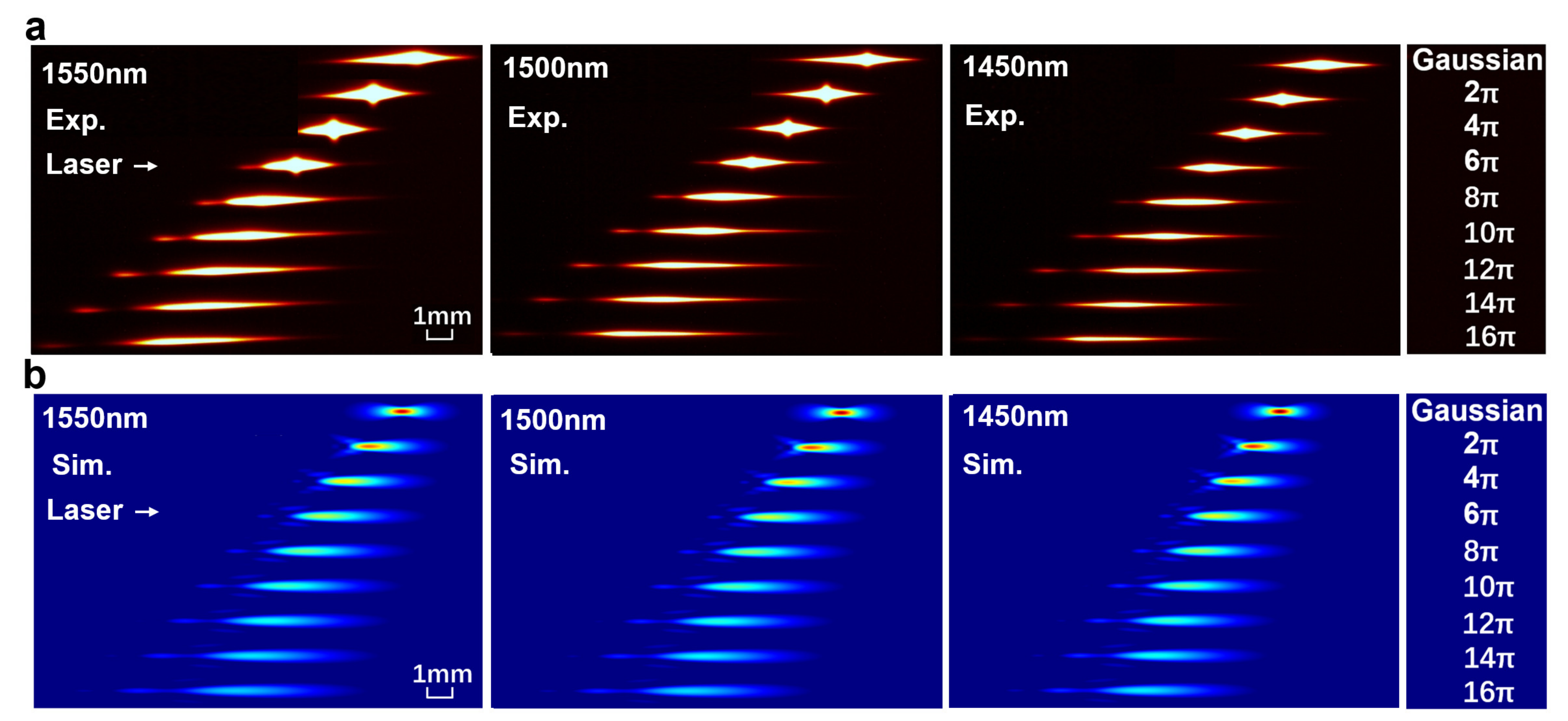, width=1.0\columnwidth}}
	\caption{\textbf{$|$ Two-color laser induced air filamentation.} \textbf{a}, Experimentally recorded fluorescence images of air-plasma filaments formed by the Gaussian and AAF beams with the increasing $C$ from $2\pi$ to $16\pi$. \textbf{b}, Laser-intensity distributions calculated by the Fresnel-diffraction theory. Here, different fundamental-laser wavelengths are taken in each column and the laser power is fixed as 100 mW.\label{Fig3}}
\end{figure}

\textbf{Lessened dephasing length due to Gaussian Gouy phase.}
These observations cannot be explained by the conventional formula of dephasing length \cite{ref22,ref23}. The dephasing lengths of the two-color Gaussian laser beams in air are $\sim$16 mm (see Tab. \ref{table1}), which is several times longer than the Gaussian-laser-induced filament lengths in our experiments (see Fig. \ref{Fig3}). Since the filament is much shorter than the dephasing length, the dephasing effect should be ignorable \cite{ref22} and the intensity dip in the THz propagation axis should not appear. The contradiction between the theory and our experiments is because the Gouy phase of Gaussian beam is not considered. We will show that when the Gouy phase shift between the two-color beams is counted, the dephasing length will plunge to $\sim$0.6 mm (Tab. \ref{table1}), shorter than the observed filament lengths, which can well explain the appearance of dips in our experiments.

\begin{table}[!htbp]
	\centering
	\caption{\textbf{$|$ Calculated dephasing lengths.} Dephasing lengths $L_{PW}$ (plane-wave approximation), $L_{Gouy}$ (with Gouy phase), $L_{AAF}$ (AAF beam) versus the fundamental-laser wavelength $\lambda_{\omega}$  (first column), where the incident laser spot radius 4 mm and the lens focal length 100 mm are taken. To calculate $L_{PW}$ including the influences of air and plasma, we adopt the plasma density corresponding to the plasma frequency of 1 THz, around the THz peak frequency observed in our experiments. \label{table1}}
	\begin{tabular}
		{cccc}
		\hline\hline
		$\lambda_{\omega}$ & $L_{PW}$ & $L_{Gouy}$ & $L_{AAF}(C=14\pi)$ \\
		\hline
		1550 nm &  16.20 mm  &  0.62 mm &  2.48 mm\\
		1500 nm &  16.35 mm  &  0.60 mm &  2.39 mm\\
		1450 nm &  16.38 mm  &  0.58 mm &  2.30 mm\\
		\hline\hline
	\end{tabular}
\end{table}

A dephasing length can be defined as $c \lambda_{2\omega}/(4\Delta v_{p})$, where $c$ is the light speed in vacuum, $\Delta v_{p}$ is the phase-velocity difference between the fundamental laser and its second harmonic, and $\lambda_{2\omega}$ is the wavelength of the latter. With plane-wave approximation, the dephasing length is
\begin{equation}\label{eq2}
	L_{PW}=\frac{\lambda_{2\omega}}{4(n_{2\omega}-n_{\omega})},
\end{equation}
where $n_{\omega}$ and $n_{2\omega}$ are the refractive indices of the fundamental and second-harmonic beams in air filament. When the beams propagate over $L_{PW}$, their phase shift reaches $\pi/2$. When Gouy phases of Gaussian beams is considered, their phases are $\varphi_{\omega}=n_{\omega}kz-\omega t-\arctan(z/z_{R})$  and $\varphi_{2\omega}=2n_{2\omega}kz-2\omega t-\arctan(z/z_{R})+\varphi_{0}$, respectively, where $\omega$ and $k$ is the fundamental-laser frequency and wavenumber, and $\varphi_{0}$ is the initial phase. By BBO frequency doubling, the Rayleigh length $z_{R}$ of the second-harmonic beam can be regard as the same with the fundamental one in  small-signal approximation \cite{ref29}. Then, their phase velocities are $v_{\omega}\approx\frac{\omega}{k}(1+\frac{1}{k}\frac{z_{R}}{z^{2}+z^{2}_{R}})$ and $v_{2\omega}\approx\frac{\omega}{k}(1+\frac{1}{2k}\frac{z_{R}}{z^{2}+z^{2}_{R}})$, respectively, presenting the phase-velocity difference $\Delta v_{Gouy}\approx\frac{\omega}{2k^{2}}\frac{z_{R}}{z^{2}+z^{2}_{R}}$ near the focal plane where most THz emission is generated. Therefore, the Gouy phase brings additional phase shift and introduces a new dephasing length:
\begin{equation}\label{eq3}
	L_{Gouy}=\frac{c\lambda_{2\omega}}{4\Delta v_{Gouy}}\approx2z_{R},
\end{equation}
where the detailed derivation is given in Sec II in SM. $L_{Gouy}$ is 26-fold smaller than $L_{PW}$, as shown in Tab. \ref{table1}. The Rayleigh length $z_{R}$, linearly proportional to the wavelength, is the value after the Gaussian beam is focused by the lens, so $L_{Gouy}$ lessens from 0.62 mm to 0.58 mm as the fundamental-laser wavelength decreases from 1550 nm to 1450 nm (see Tab. \ref{table1}), causing the dip in the angular distribution becomes deeper [see Fig. \ref{Fig2}a].

\textbf{Enhanced dephasing length with AAF beams.}
When an additional phase $\psi_{SLM}\left(R\right)=-CR^{3}/w_{SLM}^{3}=-kR^{2}/(2F_{SLM})$  is introduced, the Gaussian beam is converted to AAF one, where $F_{SLM}=\pi w_{SLM}^{3}/(C\lambda R)$ is an equivalent focal length. Since $F_{SLM}$  depends on transverse position $R$, laser intensities at different $R$ focus on different focal planes. This causes that the focusing process of AAF beam covers in a large longitudinal zone (see Fig. S3 in SM), while the focalization of Gaussian beam mainly occurs near the focal plane. Our simulations in terms of the Fresnel-diffraction theory \cite{ref30} show that the phase-velocity difference $\Delta v$ nearly linearly decreases as $C$ grows and it can be fitted by $\Delta v=\Delta v_{Gouy}-\alpha C$ with $\alpha=3.6\times 10^{-6}c$ (see Fig. S4 in SM). Then, the dephasing length can be expressed by
\begin{equation}\label{eq4}
	L_{AAF}\approx\frac{\lambda_{2\omega}}{4}\frac{c}{\Delta v_{Gouy}-\alpha C},
\end{equation}
Therefore, the phase $\psi_{SLM}\left(R\right)$ acts as a “multi-focal-length lens” and causes the phase-velocity difference between the two-color beams to reduce in the laser focusing range, i.e., increase the dephasing length. As $C$ grows to $14\pi$, the dephasing length $L_{AAF}$ climbs to $\sim$2.5 mm (Tab.\ref{table1}). This can well explain the experimental observation in Fig. \ref{Fig2} that: as $C$ increases from 0 to $14\pi$, the dip fades and eventually disappears, because $L_{AAF}$ grows to be larger than or on the order of the effective plasma-filament length (Fig. \ref{Fig3}). 

In Fig. \ref{Fig2}b we take a far-field model including THz field element interference \cite{ref25}, allowing to calculate the THz angular distribution.  In the simulations, the same laser parameters are taken as our experiments and the dephasing lengths of the Gaussian and AAF beams are adopted according to Eqs. \textcolor{blue}{(}\ref{eq3}\textcolor{blue}{)} and \textcolor{blue}{(}\ref{eq4}\textcolor{blue}{)}, respectively. The simulations agree well with the observed angular distributions shown in Fig. \ref{Fig2}a. This consistency suggests the validation of the scheme that increasing dephasing length with AAF beam can eliminate the dip.

\textbf{Improved THz emission directionality.}
Figure \ref{Fig2}c  also show that the THz emission directionality is improved or THz diameter is reduced, where we calculate the diameter $d=\sqrt{\frac{8\iint r^{2}E(r,\varphi)ds}{\iint E(r,\varphi)ds}} $ on the measured cross-sectional plane and $r$ is the distance to the centroid. This is because the filament transverse size  (THz emission source) turns smaller with growing $C$ [Fig. \ref{Fig3}a]. The reduced THz diameter with growing $C$ also shows in our far-field simulations [Fig. \ref{Fig2}b] when a thinner and longer filament is taken according to the experiments [Fig. \ref{Fig3}a], supporting that the THz diameter tends to decrease with thinner filament. With the same reason, as the plasma filament becomes thinner with the lessening laser wavelength \cite{ref31,ref32} from 1550 nm to 1450 nm [Fig. \ref{Fig3}a], the THz diameter reduces [Figs. \ref{Fig2}c and  \ref{Fig2}d]. 

Figure  \ref{Fig3}a presents the fluorescence images of two-color air plasma filaments recorded by a CCD camera. It shows that as $C$ increases, the filament becomes longer and thinner and the central intensity weakens, accompanying with the shift of filament position towards to the focusing lens. The similar results can be seen from the simulations of the laser intensity distributions exhibited in Fig.  \ref{Fig3}b, where the laser propagation through the SLM and lens is calculated according to the Fresnel-diffraction theory \cite{ref30}. The consistency between these two results is because the fluorescence intensity depends on the plasma density and temperature \cite{ref33}, which mainly determined by the laser intensity. According to our simulations presented in Fig. S3 in SM, as the transverse position of the laser field deviates from the propagation axis, the focal plane moves towards the lens for a given $C$. For a given transverse position of the laser field, the focal plane is closer to the lens with the larger $C$. This causes that the focusing process of the AAF beam covers in a large longitudinal zone and the zone range increases with $C$, corresponding to a longer plasma filament. Since the total laser power is fixed in our experiments and simulations, the longer filament tends to be thinner. Note that the filament is formed by laser-air interaction including the plasma dynamics, therefore, the filament pattern detail does not perfectly coincide with the laser-intensity distribution.

\begin{figure}[!htbp]
	\centering{\epsfig{figure=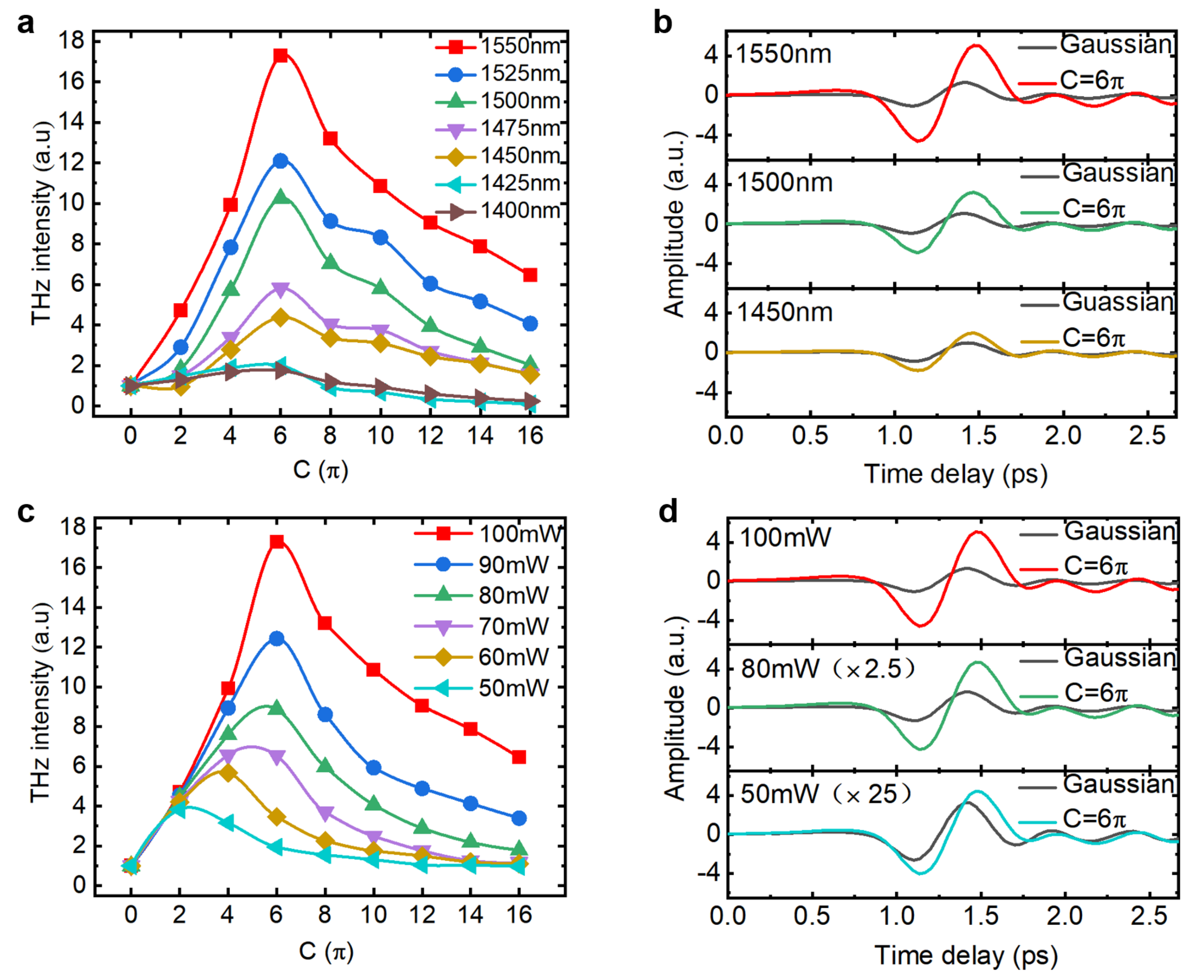, width=1.0\columnwidth}}
	\caption{\textbf{$|$ Observed THz waveform and intensity.} \textbf{a}, \textbf{c}, THz intensity versus $C$, where the intensity is normalized to the one with the Gaussian beams. In \textbf{a}, the laser power fixed as 100 mW and the fundamental-laser wavelengths ranging from 1400 nm to 1550 nm are taken, as shown by different curves. In \textbf{c}, the fundamental-laser wavelength fixed as 1550 nm and the laser powers ranging from 50 mW to 100 mW are taken, as shown by different curves. \textbf{b}, \textbf{d} Typical THz pulse waveforms with the AAF beams of $C=6\pi$ and the Gaussian beams (black curve), where each plot corresponds to different fundamental-laser wavelengths in \textbf{b} and different laser powers in \textbf{d}.\label{Fig4}}
\end{figure}

\textbf{Enhanced THz intensity.}
Figures  \ref{Fig4}a and \ref{Fig4}c illustrate that when the Gaussian beams are replaced by the AAF ones, THz intensity is enhanced, holding at various laser wavelengths and powers. In the two figures, THz intensity firstly enhances, then declines with growing $C$, and an optimized $C$ appears at a moderate value. When the laser power fixes at 100mW, the optimized $C$ appears around $6\pi$ for different laser wavelengths [see Figs. \ref{Fig4}a and \ref{Fig4}b]. This is because the dephasing length $L_{AAF}$ increases with growing $C$ [see Eq. \textcolor{blue}{(}\ref{eq4}\textcolor{blue}{)} and Fig. S4 in SM], more THz-source elements emitted from the plasma filament have the same polarity, and coherent superposition of the elements significantly improves overall THz intensity. Besides, the filament becomes longer with growing $C$ [Fig. \ref{Fig3}a], which brings more THz-source elements and tends to enhance THz intensity. On the other hand, the laser intensity peak [Fig. \ref{Fig3}b] and filament-plasma density (or ionization rate) [Fig. \ref{Fig3}a] decreases with growing $C$, which causes reduction of THz intensity. Competition between the increased $L_{AAF}$ and decreased ionization causes an optimized $C$ for the THz intensity. Note that $L_{AAF}$ nearly does not vary with the laser wavelength (Fig. S5 in SM) and the filament-plasma density (or ionization rate) is mainly determined by the laser intensity rather than its wavelength \cite{ref34,ref35,ref36}, so the optimized $C$ is independent of the laser wavelength [Fig.  \ref{Fig4}a].

While the laser power decreases, the optimized $C$ lessens from $6\pi$, as seen in Fig. \ref{Fig4}c. First, $L_{AAF}$ is independent of the laser power. Second, Fig. S6 in SM with different laser powers illustrate that the laser intensity and ionization decrease with growing $C$. When the laser power is lower and the ionization is relatively low, increasing $C$ causes the ionization to fall more significantly. Therefore, a lower laser power corresponds to a smaller optimized $C$.

THz intensity enhancement strongly depends on the laser power and wavelength, where the enhancement reaches 17-fold at $C=6\pi$ when 100 mW power and 1550 nm wavelength are taken, as shown in Figs. \ref{Fig4}a and  \ref{Fig4}c (5.3-fold enhancement was observed with ring-Airy lasers in Ref. \cite{ref26}). In our adopted parameter range, the ionization rate sharply increases with the laser power, therefore, the THz intensity enhances with the laser power. Our previous study revealed the THz strength linearly proportional to the laser wavelength \cite{ref37}. Long-distance laser nonlinear propagation can even cause the THz strength to nonlinearly grow with the wavelength \cite{ref19,ref38}. These previous results can explain our observation that the THz intensity increases with the laser wavelength.

\textbf{Conclusion.}
In summary, we have proposed a scheme to efficiently suppress the dephasing effect in the two-color-laser-driven THz generation in air, by replacing ordinary Gaussian laser beams with AAF ones. The AAF beam is easily achieved by imposing an additional phase of $-CR^{3}/w_{SLM}^{3}$ on the input Gaussian beam. This phase acts as a “multi-focal-length lens” and causes the two-color laser phase-velocity difference to reduce, i.e., increase dephasing length. When the dephasing length is larger than efficient air-filament length, our experiments and simulations show that the dip in the THz angular profile can be eliminated. The “multi-focal-length lens” can also increase the filament length, which combining with the extended dephasing length can improve the THz emission directionality and enhance the THz intensity by 17 folds.

\begin{methods}
\textbf{THz pulse detection setup\\}
The experimental setup is depicted in Fig. S1\textbf{a}. The horizontally polarized Gaussian laser beam with wavelength from 1200 nm to 1600 nm is introduced into the SLM loaded with AAF phase distribution. The laser beam is focused by a plano-convex lens with a focal length of 100 mm. The BBO crystal is placed between the lens and plasma to create the second-harmonic beam. The THz wave generated from the two-color laser induced air plasma is collimated and refocused by a pair of off-axis parabolic mirrors (PM) after eliminating the residual femtosecond laser beam by a silicon wafer and low-pass filters. The 800 nm probe beam after a delay line (not shown in the figure) is focused by a plano-convex lens and then passes through a small hole on the back of second parabolic mirror. The colinearly propagated THz wave and probe beam are focused onto a 1-mm-thick $<110>$ cut ZnTe crystal. The THz time-domain waveform is recorded by electro-optic sampling using a detection device combined by the quarter waveplate (QWP), Wollaston prism (WP) and balanced detector (BD). Fig. S1\textbf{b}. shows the AAF phase distributions loaded on the phase mask of SLM with several representative modulation depths.

\noindent\textbf{Fresnel diffraction theory\\}
Using the Fresnel diffraction theory \cite{ref30}, we calculate the phase velocity differences between the two-color AAF beams,  and the corresponding dephasing lengths. The Fresnel diffraction theory allow us calculate the electric field strength distribution $E(r,z)$ at the longitudinal position $z$ according to its initial transverse electric field strength distribution at the positon $z_{0}$:
\begin{equation}\label{eq5}
  E\left(r,z\right)=-\frac{2\pi i}{\lambda \left(z-z_{0}\right)}e^{i\frac{\pi}{\lambda (z-z_{0})}r^{2}}\int_{0}^{\infty}E\left(r^{\prime},z_{0}\right)J_{0}\left[\frac{2\pi rr^{\prime}}{\lambda \left(z-z_{0}\right)}\right]e^{i\frac{\pi}{\lambda (z-z_{0})}r^{\prime2}}r^{\prime}\mathrm{d}r^{\prime}
\end{equation}
where $J_{0}$ is the zero-order Bessel function, $r$ is the transverse coordinate, and $i$ is the imaginary symbol. The SLM and the lens act as phase diffraction elements which can impose phase $\psi_{SLM}$ and $\psi_{F}=-\frac{\pi r^{2}}{\lambda F}$ on $E(r,z)$, where $F=100$ mm in our experiment. When the laser propagates through the BBO crystal, its frequency $\omega$ becomes $2\omega$ and the new field strength is proportional to $E^{2}$. In this way, we also calculate the laser intensity distributions shown in Fig. \ref{Fig3}.
\end{methods}

\begin{addendum}
\item [Acknowledgements] This work was supported by the National Natural Science Foundation of China (Grant Nos. 12074272, 11775302, 61905271), the Strategic Priority Research Program of Chinese Academy of Sciences (Grant No. XDA25050300), the R\&D Program of Beijing Municipal Education Commission (Grant No. 23JD0035), the National Key R\&D Program of China (Grant No. 2018YFA0404801), the Youth Beijing Scholar Program of the Beijing Government, the Guangdong Basic and Applied Basic Research Foundation (Grant No. 2023A1515011397), the Fundamental Research Funds for the Central Universities, the Research Funds of Renmin University of China (20XNLG01). Computational resources have been provided by the Physical Laboratory of High Performance Computing at Renmin University of China. A portion of this work was carried out at the Synergetic Extreme Condition User Facility (SECUF).

\item[Author contributions] 
W.-M.W. and L.-L.Z. conceived the main idea and wrote the main manuscript. X.-R.Z. and L.-L.Z. performed the experiment. N.L. and W.-M.W. conducted theoretical model and simulation. X.-R.Z. and N.L. prepared the draft with the supervision of L.-L.Z. and W.-M.W. All authors commented on the manuscript.

\textbf{Competing interests} Authors declare no competing interests.

\textbf{Supplementary Information} Supplementary Materials (SM) accompanies this paper.\\
\textbf{Reprints and permission information} is available online.

\end{addendum}
\nolinenumbers

\section*{References}

\end{document}